# Use of Effective Audio in E-learning Courseware

Kisor Ray
Techno India Agartala, an Engineering College affiliated to Tripura University,
Maheshkhola, Tripura (West), India

*Abstract— E-Learning uses electronic media, information & communication technologies to provide education to the masses. E-learning deliver hypertext, text, audio, images, animation and videos using desktop standalone computer, local area network based intranet and internet based contents. While producing an e-learning content or course-ware, a major decision making factor is whether to use audio for the benefit of the end users. Generally, three types of audio can be used in e-learning: narration, music and sound effect. This paper shows that the use of proper audio based on contents type and subject can make the content more interesting as well as help the end users to better understand the contents.*

## I. INTRODUCTION

E-learning is a learning tool which can enhance the capability of a learner to understand a specific subject easily. However, acceptability of such tool depends on how much interest it can generate among the end users. So, the design of such tool at implementation level should be carefully examined at its each component level so that depending on situations a flexible design should allow different kind of implementations for each component. Audio is an important component of e-learning courseware. Among the three types of audio used in e-learning courseware, narration could be of four sub-types like Elaborative (on-screen text summarizes the audio), Paraphrasing (audio summarizes the on-screen text), Verbatim (reading exact words on-screen) and Descriptive (audio describes image on-screen) [1].The goal of this paper is to investigate on the use of audio for e-learning courseware and to conclude about the types of audio that may be used to produce quality e-learning courseware which can have better acceptability by generating interest about the subjects it's dealing with, thus producing better performance for the end users.

## II. MOTIVATION

Interest of the students is the key for the success of e-learning courseware. The greater the interest, more involvement and more understanding about the subjects. The success or failure of a courseware is related to student motivation. An interesting courseware is more likely to generate better motivation. A better motivated student is more likely to put better effort thus producing better performances. So, each of the important components of an e-learning courseware needs to be examined carefully at the implementation level. This paper intends to examine the varied implementation of different audio types and their effects on the performance level of the participating students in order to suggest certain implementation styles while designing an e-learning courseware.

## III. METHODS, PROCEDURE AND THE PARTICIPANTS

To investigate the effect on the use of audio in the e-learning courseware, we selected three subjects : Physics, Chemistry and Mathematics for an age group of 13+ covering ICSE (Council for the Indian School Certificate Examinations) syllabus for VII standard. The participants for the experimental part were the class VII standard students from English medium schools under the ICSE board. Many students find ICSE syllabus vast as compared to CBSE (Central Board of Secondary Education) for its detailed and elaborated coverage from the early stage of middle school like class (grade) VII. Each courseware was designed to explain the fundamental concept of each topic selected, followed by the summarized animated texts with an objective to develop different skills like psychomotor, interpersonal and cognitive [2]. Each courseware included learning resources, interactive lessons, simulations and job aids . Use of numerous graphics and animations based on Flash and HTML 5.0 were the major features of the courseware. Each conceptual unit was followed by a short question answer review section. Eight to ten conceptual units formed a courseware for a particular topic. At the end of all the units, every topic wise courseware contained a question bank of 40-60 multiple choice questions (MCQs) for the evaluation of each student participated to gather knowledge and their understanding from the selected topic. Courseware on the selected topics from each subject was prepared using different types of audio with a varied implementation styles (Table I, II and III) with an aim to gather relevant data in a systematic and orderly manner for the purpose of investigation.

TABLE I
COURSEWARE ON DIFFERENT PHYSICS TOPICS PREPARED WITH TYPES OF AUDIO

| Subject | Topic | Audio Types |
|---|---|---|
| Physics | Motion | Elaborative |
|  | Heat | Paraphrasing |
|  | Light | Verbatim |
|  | Sound | Descriptive |

TABLE II
COURSEWARE ON DIFFERENT CHEMISTRY TOPICS PREPARED WITH TYPES OF AUDIO

| Subject | Topic | Audio Types |
|---|---|---|
| Chemistry | Physical & Chemical Change | Elaborative |
|  | Atomic Structure | Paraphrasing |
|  | Composition of Air | Verbatim |
|  | Elements & Compound | Descriptive |

TABLE III
COURSEWARE ON DIFFERENT MATHEMATICS TOPICS PREPARED WITH TYPES OF AUDIO

| Subject | Topic | Audio Types |
|---|---|---|
| Mathematics | Unitary Method | Elaborative |
|  | Number System | Paraphrasing |
|  | Basic Algebra | Verbatim |
|  | Set Theory | Descriptive |

The audio used was generated using text to speech (TTS) technology. Effort was given to produce near natural sounding voices as synthesized output using customized voice xml (VXML) with Indian accent. Attention was given to pauses, non-lexical phenomena, utterances, kinesics and changes in vocal quality [3] while making the audio transcripts for the 'text to speech engine' as input. Adobe Captivate™ , Articulate Storyline™ and eXe tools etc. were used to generate modify and customize the HTML and SWF contents along with the use of java script and custom triggers. Animations were created using Microsoft Power Point™, Adobe Macromedia™, Toon Boom™ etc. Mass production of quizzes (MCQs) were done in GIFT format so that they could be easily imported into the captivate platform. The Courseware for each subject was designed to deliver contents in an instructorless environment.

IV. RESEARCH HYPOTHESIS

In this study, five research hypotheses were examined to determine which hypothesis should be accepted and which should be rejected.
H1: Generally, students have preference on types of audio used for Courseware
H2: Specific preferences of student on the types of audio used for specific subjects
H3: Students' performance in the evaluation test will reflect the types of audio used for courseware
H4: Effect of types of audio used may differ from subject to subject
H5: Courseware effectiveness does not depend on types of audio used

V. STUDIES AND EVALUATION

The study mentioned in this paper investigated H1, H2, H3, H4 and H5. Students were asked to undertake lessons using courseware for each subject (Physics, Chemistry and Mathematics) in an instructorless environment. After completion of each topic of the respective subjects, they had taken the evaluation test and a survey about their likings and interest level were also conducted. In the Table IV,V, and VI the outcome of the survey is given on column 4 as 'level of interest'. This column actually reflects the 'likings and interest' of the majority of the participants. This study obtained various data shown in the Table IV, V, VI, VII, VIII, and IX to establishing certain correlation between the various audio types implementation in courseware vis-à-vis performances of the participants.

TABLE IV
EFFECT OF AUDIO TYPES USED IN PHYSICS COURSEWARE

| Subject | Topic | Audio Type | Interest Level | Evaluation Test Performance |
|---|---|---|---|---|
| 20 | Motion | Elaborative | Medium | Low |
| 20 | Heat | Paraphrasing | High | High |
| 20 | Light | Verbatim | Low | Low |
| 20 | Sound | Descriptive | High | Medium |

TABLE V
EFFECT OF AUDIO TYPES USED IN CHEMISTRY COURSEWARE

| Subject | Topic | Audio Type | Interest Level | Evaluation Test Performance |
|---|---|---|---|---|
| Chemistry | Physical & Chemical Change | Elaborative | Medium | Low |
| | Atomic Structure | Paraphrasing | High | High |
| | Composition of Air | Verbatim | Low | Low |
| | Elements & Compound | Descriptive | High | Medium |

TABLE VI
EFFECT OF AUDIO TYPES USED IN MATHEMATICS COURSEWARE

| Subject | Topic | Audio Type | Interest Level | Evaluation Test Performance |
|---|---|---|---|---|
| Mathematics | Unitary Method | Elaborative | Medium | Low |
| | Number System | Paraphrasing | High | High |
| | Basic Algebra | Verbatim | Low | Low |
| | Set Theory | Descriptive | High | Medium |

TABLE VII
SCORE OBTAINED BY THE STUDENTS IN DIFFERENT TOPICS OF PHYSICS (P1,P2 ,P3 AND P4) UNDER VARIED AUDIO TYPES

| No. of Students | Topic | Audio Type | Average Score in Evaluation Test |
|---|---|---|---|
| Physics | Motion | Elaborative | 80% |
| | Heat | Paraphrasing | 94% |
| | Light | Verbatim | 78% |
| | Sound | Descriptive | 89% |

TABLE VIII
SCORE OBTAINED BY THE STUDENTS IN DIFFERENT TOPICS OF CHEMISTRY (C1,C2 ,C3 AND C4) UNDER VARIED AUDIO TYPES

| No. of Students | Topic | Audio Type | Average Score in Evaluation Test |
|---|---|---|---|
| Chemistry | Physical & Chemical Change | Elaborative | 79% |
| | Atomic Structure | Paraphrasing | 96% |
| | Composition of Air | Verbatim | 74% |
| | Elements & Compound | Descriptive | 90% |

TABLE IX
SCORE OBTAINED BY THE STUDENTS IN DIFFERENT TOPICS OF MATHEMATICS (M1,M2 ,M3 AND M4) UNDER VARIED AUDIO TYPES

| No. of Students | Topic | Audio Type | Average Score in Evaluation Test |
|---|---|---|---|
| Mathematics | Unitary Method | Elaborative | 78% |
| | Number System | Paraphrasing | 96% |
| | Basic Algebra | Verbatim | 81% |
| | Set Theory | Descriptive | 97% |

## VI. DISCUSSION AND FINDINGS

We have a number of sets here namely:

(i) ST = {1,2,3,……20} , a set of 20 students ,

(ii) SJ = {P,C,M } , a set of three subjects Physics, Chemistry and Mathematics

(iii) CW = {P1,P2,P3,P4,C1,C2,C3,C4,M1,M2,M3,M4} , a set of 12 course-ware

(iv) AT = { Narration, Music ,Sound effect} , A set of audio types

(v) AST = {Elaborative, Paraphrasing, Verbatim, Descriptive} , a set of audio sub-type from the main type 'narration'.

(vi) LVL = {High, Medium ,Low} , a set of levels for the test score or interest

Audio type 'narration' and its 04 subtypes (set AST) were mainly used in the 12 course-ware ( set CW) whereas audio types namely 'music' and 'sound effects' ( part of set AT) were used mainly in quizzes only. What we find that sets like

SJ, CW, AST and LVL have close correlation. Subjects like physics and chemistry become more interesting when the audio sub type 'Paraphrasing' is used at the end of each slide and/or page summarizing the main concept as a recapitulation after a participant completed the study of each slide in a courseware, whereas mathematics becomes more interesting when the audio sub type 'Descriptive' is used like a 'voice of a virtual teacher explaining' after each line of the animated mathematical expression/equation appears on the slide. Application of audio sub types 'Elaborative' and 'Verbatim' is found to generate lower interest among the participants irrespective of the subjects thus resulting lesser performances for the courseware where these two were implemented.

## VII. CONCLUSION AND FURTHER WORK

From our study what we can conclude in this paper is that the effectiveness of e-learning courseware depends on the implementation of its key components. Audio is an important component of e-learning courseware. Implementation of varied types of audio in the courseware has its own effect. We also observe that (1) the participants have their preferences on the types of audio used in the courseware supporting H1. (2) Preferences on the types of audio varies with subject types. For example, while mathematics courseware is found suitable for the implementation of 'Describe' sub type of audio, physics and chemistry are found to be suitable for the implementation of the 'Paraphrasing' sub type of audio, thus supporting H2 (3). Performance of a participant depends on the interest as well as motivation level, should the participant like the implementation style of the audio, his/her interest and motivation level will be higher producing better performance. What we can see that for physics and chemistry 'Paraphrasing' and for mathematics 'Describe' sub types of audio are liked by the participant producing better performance in the respective evaluation tests, which supports H3. (4) We have also seen that the influence/effect of the type of audio is not same for every subject. What is good for subjects like physics or chemistry may not be as good for subjects like mathematics, which supports H4. (5) We also find that the effectiveness of the courseware depends on the types of audio used since we have already seen that audio types 'Elaborative' and 'Verbatim' are less influential in producing better performance thus contradicting H5. In order to provide more support to our findings, further study & investigation may be required. An alternative study as defined below may be followed to collect systematic data to analyses and compare with our findings. The participants may be divided into 04 different sets (groups) that is instead of considering all the 20 students/participants a single set ST, like :

ST1{1,2,..5} - a set of 05 students/participants
ST2 {6,7,..10} - a set of 05 students/participants
ST3 {11,12,..15} - a set of 05 students/participants
ST4 {15,16,..20} - a set of 05 students/participants

Let the courseware group type be on the basis of audio sub types that is the set of the sets of all the courseware based on the specific audio sub types, for example:

CWTe={CWPe,CWCe,CWMe} - a family of sets of courseware type based on audio sub types 'Elaborative'
CWTd={CWPd,CWCd,CWMd} - a family of sets of courseware type based on audio sub types 'Describe'
CWTp={CWPp,CWCp,CWMp} - a family of sets of courseware type based on audio sub types 'Paraphrasing'
CWTv={CWPv,CWCv,CWMv} - a family of sets of courseware type based on audio sub types 'Verbatim'
STS={ST1,ST2,ST3,ST4} - a family of sets of students based on all the student groups. So, the alternative method needs to create 48 such course wares because each subject needs to cover four topics under each audio sub type. Each of the family of sets of courseware type based on audio sub types may be applied to the family of sets consisting all the students/participants and average performance that could be computed.
STSPSe = CWTe->STS is the average performance score of the all the student groups undertaking all courseware type produced using audio subtype 'Elaborative'.
STSPSd = CWTd->STS is the average performance score of the all the student groups undertaking all courseware type produced using audio subtype 'Describe'.
STSPSp= CWTd->STS is the average performance score of the all the student groups undertaking all courseware type produced using audio subtype 'Paraphrasing'.
STSPSv= CWTd->STS is the average performance score of the all the student groups undertaking all courseware type produced using audio subtype 'verbatim'. In this suggested alternative method, one can have all the average performance scores which may be used for the comparison purpose. In our study, we have shown more inclination towards the audio type 'narration' and its sub types 'Elaborative', 'Describe', 'Paraphrasing' and 'Verbatim' with lesser attention to 'music' and 'sound effect'. In our courseware, especially for physics and chemistry as and when required we have used the sound effect irrespective of the narrative audio ('Elaborative', 'Describe', 'Paraphrasing' and 'Verbatim' ) to explain the subjects with animation and no direct/in-direct measure is instituted to find the influence of the 'sound effect' separately. Music and audio were also used for quizzes with a finding that the participants are having lower interest in them especially in 'quiz/test' type of situation. We also restricted ourselves from using the same courseware with different audio sub types ( Eg. Creation of courseware with topic 'Motion' under the subject physics using all four subtypes of audio) to undertake the study of the effect of the types of audio on the same group of participants presuming that cumulative knowledge gathered on that specific subject by the end users through repeated use may actually prevent us from collecting the effective data .